# Accelerated Molecular Vibrational Decay and Suppressed Electronic Nonlinearities in Plasmonic Cavities through Coherent Raman Scattering

Lukas A. Jakob[1†], William M. Deacon[1†], Rakesh Arul[1], Bart de Nijs[1], Niclas S. Mueller[1], Jeremy J. Baumberg[1*]

[1] Nanophotonics Centre, Cavendish Laboratory, University of Cambridge, Cambridge CB3 0HE, UK.
[†] Authors contributed equally; [*] jjb12@cam.ac.uk

KEYWORDS: coherent Raman scattering, SERS, vibrational dynamics, time-resolved, plasmonics

## Abstract

**Molecular vibrations and their dynamics are of outstanding importance for electronic and thermal transport in nanoscale devices as well as for molecular catalysis. The vibrational dynamics of <100 molecules are studied through three-colour time-resolved coherent anti-Stokes Raman spectroscopy (trCARS) using plasmonic nanoantennas. This isolates molecular signals from four-wave mixing (FWM), while using exceptionally low nanowatt powers to avoid molecular damage via single-photon lock-in detection. FWM is found to be strongly suppressed in nm-wide plasmonic gaps compared to plasmonic nanoparticles. The ultrafast vibrational decay rates of biphenyl-4-thiol molecules are accelerated ten-fold by a transient rise in local non-equilibrium temperature excited by enhanced, pulsed optical fields within these plasmonic nanocavities. Separating the contributions of vibrational population decay and dephasing carefully explores the vibrational decay channels of these tightly confined molecules. Such extreme plasmonic enhancement within nanogaps opens up prospects for measuring single-molecule vibrationally-coupled dynamics and diverse molecular optomechanics phenomena.**

## I. Introduction

The intense optical fields elicited by plasmonic nanostructures allow routine access to spectra of single/few-molecules under ambient conditions [1–3]. This starts to open prospects for understanding dynamics and chemistry of individual molecules in real time. Using high-speed imaging, it is possible to track the dynamics of few molecules on the microsecond scale [4] using surface-enhanced Raman spectroscopy (SERS). However, pushing the time resolution down to the scale of molecular vibrations (picoseconds) with ultrafast time-resolved spectroscopies has proven extremely difficult [5,6]. Early measurements of molecular phonon dynamics through surface-enhanced coherent anti-Stokes Raman spectroscopy (SE-CARS) on single nanostructures suggested great promise [7–9], however since then little progress has been made [5,10]. Molecular dynamics studied recently in ultrafast pump-probe spectroscopy of spontaneous SERS instead averages over many molecules [11,12].

Single- and few-molecule SE-CARS and SERS measurements are possible only due to the high field intensities and Raman enhancements achieved when exciting plasmonic cavities with ultrafast pulses [13]. However, an unwanted consequence of these extreme peak fields is a rapid increase in nanoscale damage compared to continuous wave measurements [5,14–16], which modifies both the metallic and molecular components of the system. This is particularly a problem for electronically-resonant molecules which enhance signals, but are damaged much more easily. As a result, time-resolved SE-CARS (trCARS) and time-resolved spontaneous SERS (trSERS) measurements which demand the structure be stable throughout, become impractical. Although the damage thresholds are known, the mechanisms behind this damage remain unresolved [17], being most likely due to non-equilibrium heating and optical forces acting directly on the surface gold atoms [15,18].

The second problem with SE-CARS is the large overlapping four-wave mixing (FWM) signal due to electronic FWM [19–21]. Claims that the short FWM lifetime enables separation of this from SE-CARS assume that the molecular vibrational lifetime is much longer than the electronic lifetime which, as we demonstrate below, is not the case for molecules near metal surfaces.

To circumvent these problems, we develop a three-colour CARS scheme to measure the reduction in phonon lifetimes of exemplar biphenyl-4-thiol molecules (and other molecules, see Supplementary Information) in tightly-confined plasmonic cavities. Surprisingly, the electronic FWM contribution is now found to be small ($< 10\%$) despite the penetration of light into the metal, enabling extraction of SE-CARS signatures from ~100 molecules. Further, using a newly developed single-photon timestamping technique [22], we can simultaneously measure coherent trCARS [Fig. 1(a)] and spontaneous trSERS [Fig. 1(b)] on individual nanostructures at sub-µW powers to independently probe the coherence ($T_2$) and population lifetimes ($T_1$), respectively [23]. This novel, ultrasensitive three-color CARS scheme thus allows us to independently determine the population decay and dephasing in few-molecule plasmonically-enhanced systems, giving insight into phonon dynamics of molecules tightly coupled to metallic interfaces.

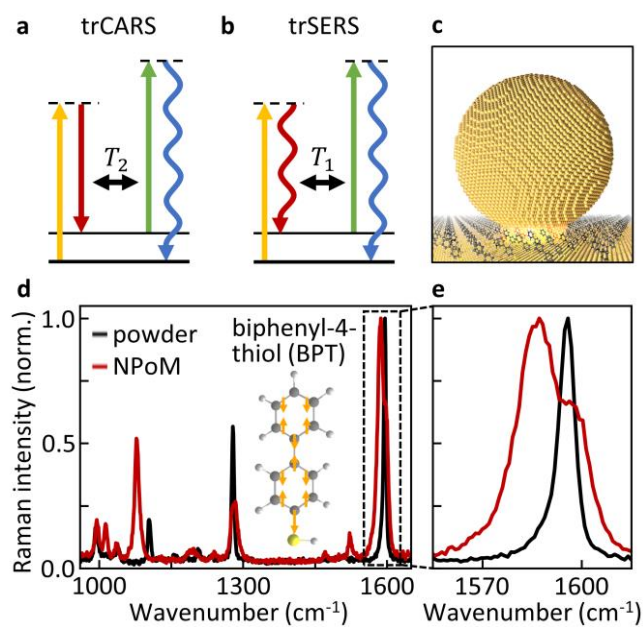

**Fig. 1 | Raman scattering from biphenyl-4-thiol (BPT) in plasmonic nanocavities. a,b**, Excitation scheme of (a) coherent anti-Stokes Raman scattering (trCARS) and (b) spontaneous pump-probe SERS (trSERS). Spontaneous (wavy arrows) vs stimulated (straight arrows) processes. **c,** Nanoparticle-on-mirror (NPoM) geometry with molecular layer in nanogap. **d**, Continuous-wave Raman spectrum of BPT powder (black) compared with SERS from BPT in the NPoM cavity (red). Inset: BPT molecule with arrows indicating the coupled ring vibrational mode (1585 cm$^{-1}$). **e**, Detailed view of coupled ring vibration in BPT. Binding to the gold surface splits and broadens the vibrational line.

## II. Experimental Setup

NanoParticle-on-Mirror (NPoM) plasmonic nanocavities confine light to the nanometre-scale gap between a metallic nanoparticle and the underlying metal substrate [Fig. 1(c)] [24,25]. These structures are also known as metal-insulator-metal (MIM) waveguides, or particle-on-film structures. This platform for few-molecule vibrational spectroscopy traps optical fields to zeptolitre volumes, filled only by the molecules under study [26]. The NPoM geometry has facilitated explorations of many processes through SERS including the movement of single atoms, tracking nano-electrochemistry, and molecular optomechanical coupling [15,27–30].

We utilise NPoM structures built from 80 nm gold nanoparticles separated from a gold substrate by a self-assembled monolayer (SAM) of biphenyl-4-thiol (BPT). To improve the outcoupling of the CARS signal and increase sample stability, the NPoM structures are covered with a thin film of PMMA which enhances the NPoM scattering resonance in the signal range around 650 nm (see Supplementary Information Fig. S1). [31] Due to molecular binding to the gold surface atoms, the molecular SERS spectra are modified compared to their powder Raman spectrum [Fig. 1(d,e)]. Notably, the coupled ring mode at 1585 cm$^{-1}$ [see inset in Fig. 1(d)] splits into two modes with separation 13 cm$^{-1}$ and is broadened in the NPoM structure compared to the powder spectrum. However, this effect is only observable with narrow-linewidth continuous-wave lasers since picosecond-pulsed lasers have a larger spectral bandwidth, thus broadening the Raman peak linewidth.

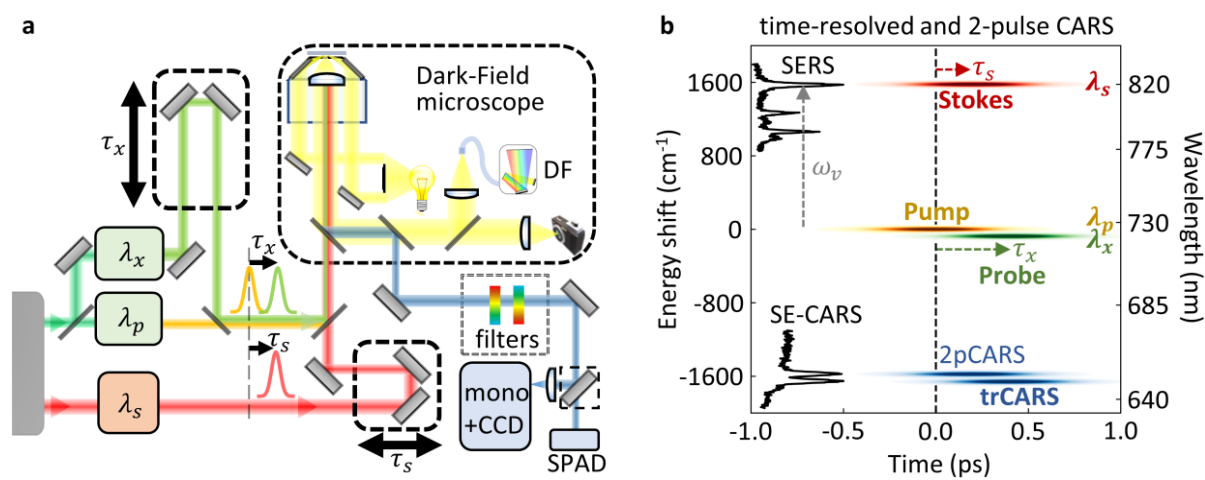

**Fig. 2 | Experimental setup. a,** Automated darkfield microscope combines three laser beams with variable time delay for SE-CARS. The signal is spectrally filtered and detected with a spectrometer (monochromator + CCD) or with a single-photon avalanche diode (SPAD) for single-photon lock-in detection. **b,** Temporal-spectral pulse tunings for pump ($p$, yellow), Stokes ($S$, red), and probe ($x$, green), giving emission (blue) from two-pulse CARS (2pCARS) and three-pulse time-resolved CARS (trCARS). Top left shows BPT SERS spectrum from pump only, bottom left BPT SE-CARS spectrum from all three pulses.

Nanoparticles are automatically located and characterised using a custom-built reflective dark-field microscope, prior to each trCARS measurement [Fig. 2(a)]. In our implementation, three temporally and spectrally independent input beams are required in the pulse train: the pump (set as $t = 0$, $\lambda_p$), the Stokes ($t = \tau_S$, $\lambda_S$), and the probe pulses ($t = \tau_x$, $\lambda_x$) [Fig. 2(b)]. The spectral shift between the pump and the Stokes pulse is chosen to populate the vibrational mode under study with frequency $\omega_v$ (here $\omega_p - \omega_S = \omega_v$ = 1585 cm$^{-1}$). To achieve this, a tuneable pulse from an optical parametric oscillator (OPO) provides $\lambda_p$ and $\lambda_x$ while the 820 nm pump laser provides $\lambda_S$. Bandpass filtering sets all pulses to 0.5 ps pulse width (1.5 nm spectral width) to balance spectral and temporal resolution and reduce damage. The spectrum of $\lambda_{p,x}$ is selected from the 10 nm-wide OPO pulse by independently-tuned bandpass filters. Pulse energies are kept low enough to avoid damage to the NPoM constructs (average powers <4 µW µm$^{-2}$, pulse energies <50 fJ µm$^{-2}$). After filtering out the laser pulses from the emitted light, signals are detected by either a spectrometer or by a single-photon detector for single-photon lock-in detection with improved signal-to-noise ratio (which is described in detail elsewhere [22]).

## III. Suppression of Electronic Four-Wave Mixing

The three-pulse scheme for SE-CARS allows different nonlinear contributions to be distinguished. Three distinct nonlinear processes can occur: two-pulse CARS ($\omega_{\text{2pCARS}}$), three-pulse trCARS ($\omega_{\text{trCARS}}$) and FWM ($\omega_{\text{FWM}}$) [Fig. 3(b)], with frequencies

$$\omega_{\text{2pCARS}} = \omega_p - \omega_S + \omega_p = \omega_p + \omega_\nu$$
$$\omega_{\text{trCARS}} = \omega_p - \omega_S + \omega_x = \omega_x + \omega_\nu$$
$$\omega_{\text{FWM}} = \omega_1 + \omega_2 - \omega_3$$

when assuming that the CARS process is on-resonance, hence $\omega_\nu = \omega_p - \omega_S$. Any of the photons involved in CARS can also interact via FWM such that $\omega_{1,2,3}$ can be equal to $\omega_{p,S,x}$. With degenerate pump and probe pulses used to perform trCARS ($\omega_p = \omega_x$), spectral separation of the 2-pulse and trCARS signals is inherently impossible as $\omega_{\text{2pCARS}} = \omega_{\text{trCARS}}$. This adds a varying background in trCARS signals. Note that phase matching is not relevant for these experiments on single sub-wavelength-sized NPoMs.

To spectrally separate 2-pulse and trCARS, the probe wavelength $\lambda_x$ must be significantly detuned from the pump wavelength $\lambda_p$. While often achieved by frequency-doubling of the pump or Stokes, this unfortunately directly pumps Au interband transitions, considerably increasing nanoscale damage. Other approaches using nonlinearly-generated or electronically-synchronised non-degenerate pulses reduce timing resolution (>1 ps), exceeding the phonon lifetime. The approach here retains spectral and temporal resolution by spectrally filtering the 10 nm-wide OPO output to create two 1.5 nm pulses for the pump and probe beams. A similar approach has recently been employed for biological imaging using stimulated Raman scattering [32].

The three-colour CARS scheme allows the BPT-filled NPoM to be driven at $\lambda_p$ = 726 nm, $\lambda_S$ = 820 nm and $\lambda_x$ = 722 nm, generating nonlinear emission spectra (Fig. 3a, with $\tau_S = \tau_x$ = 0 ps, average laser intensity 4 µW µm$^{-2}$). Two strong peaks are observed at $\omega_{\text{2pCARS}}$ = 652 nm and $\omega_{\text{trCARS}}$ = 649 nm (slightly chirped). However, the third peak at $\omega_{\text{FWM}}$ = 646 nm from pure FWM ($\omega_{\text{FWM}} = 2\omega_x - \omega_S$) is very weak. This non-resonant FWM contribution from molecular vibrations, electronic Raman scattering, and 2-photon gold luminescence is thus ten times smaller than the CARS signals. We further confirm that the molecular CARS signal dominates over FWM by scanning $\lambda_p$ across the vibrational resonance. Only when the resonance condition $\omega_p - \omega_S = \omega_\nu$ is satisfied does a strong SE-CARS emission appear (Supplementary Information Fig. S2).

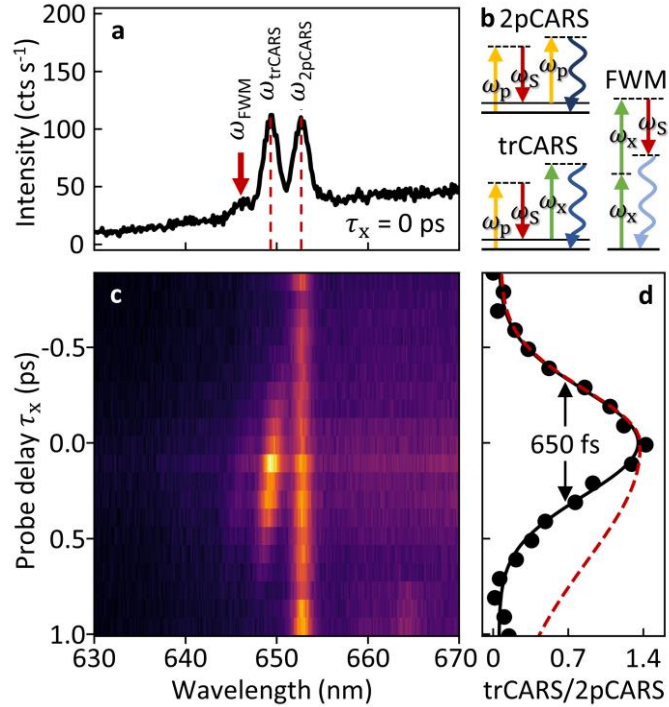

**Fig. 3 | SE-CARS spectro-temporal scan. a,** SE-CARS signal from BPT NPoM at 0 ps probe delay showing clear peaks at $\omega_{2pCARS}$ and $\omega_{trCARS}$, but minimal peak at $\omega_{FWM}$. **b,** Excitation schemes for 2pCARS (pump-Stokes-pump), trCARS (pump-Stokes-probe) and FWM (probe-probe-Stokes). **c,** Time-resolved SE-CARS spectra. **d,** Integrated trCARS signal normalised by 2pCARS signal. Black line indicates Gaussian fit with 650 fs FWHM and red dashed line is the expected vibrational decay from CW SERS. Laser intensity 4 µW.µm$^{-2}$ per beam.

These nonlinear experiments on BPT NPoMs demonstrate that the signal observed is dominated by molecular CARS. This is surprising in the context of many previous measurements that show FWM from metallic plasmonic components is very large [5,33–35]. The disparity arises from the different field distributions of different plasmonic architectures. The field $E$ in NPoMs is tightly confined to the gap (of width $d$ and permittivity $\varepsilon_g$) where the molecules reside, while field leakage into the metal (of permittivity $\varepsilon_m$) decays spatially within $\delta = d/2\varepsilon_g \, \Re\{1/\varepsilon_m\} \sim 1$ nm. [1] Although both are enhanced by the plasmonic gap, the ratio of integrated SE-CARS to FWM given by the ratio of $|E|^8$ in the gap and in the metal is thus $2|\varepsilon_m(\lambda_c)/\varepsilon_g|^7 \sim 10^6$, much larger than in other plasmonic structures (for spherical Au nanoparticles this ratio is of order unity). This also explains observations of weaker FWM in closely-spaced dimers [13]. Small-gap plasmonic structures hence generate large CARS signals (~$10^7$ counts/nJ$^3$ here, still far below saturation) and weak FWM in comparison to other plasmonic structures with 10-10$^8$–fold weaker SE-CARS [9,35,36].

## IV. Accelerated Molecular Vibrational Decay

The clear evidence that trCARS signals originate from molecular vibrations allows us to explore their phonon dynamics. Scanning the time delay of the probe pulse $\tau_x$ allows a spectro-temporal map to be built [Fig. 3(c)]. Since the 2pCARS (at 652 nm) is independent of the probe pulse, no change is expected in its intensity. However sporadic fluctuations are observed in time-scans, most likely due to nanoscale modifications leading to changes in molecular coupling efficiency [27,28,37]. This simultaneous 2pCARS allows the trCARS signal (at 649 nm) to be suitably normalised to reveal the true time-resolved intensity independent of transient signal fluctuations [Fig. 3(d)]. The extracted trCARS signal fits a Gaussian with FWHM of 650 fs corresponding to the convoluted excitation and probe pulses of 500 fs. The molecular phonon lifetime here is thus below the instrument response. In contrast, Fourier-transform analysis of CW SERS spectra [Fig. 1(d), red] suggests a molecular phonon lifetime of >600 fs for this vibrational mode, indeed detectable with this setup [Fig. 3(d), red dashed]. This is faster than the original phonon lifetime in the absence of a nearby metal surface estimated as 2 ps in BPT powder, extracted from its Raman spectrum [Fig. 1(d), black]. More detail on Fourier analysis of Raman spectra to estimate CARS time tracks can be found in Supplementary Information Fig. S3.

T This accelerated phonon decay implies that CW SERS and pulsed SE-CARS are probing the vibrational decay under different conditions. Critically, the peak powers of the pulsed lasers employed in SE-CARS are orders of magnitudes higher than the CW average power, thereby driving the entire system (both vibrational states in molecules and metal as well as electronic and plasmonic states) out of the thermal equilibrium [15]. Therefore, we reduce the pulsed laser intensity $I_l$ by 20-fold to 200 nW μm$^{-2}$ per beam (for pump, Stokes and probe lasers), corresponding to peak fields in the nanogaps of order $10^8$ V/m. We emphasise these are applied to transparent (non-electronically resonant) molecules, in contrast to other work using dye molecules which damage even more rapidly. Since trCARS signals scale $\propto I_l^3$, recorded count rates are now almost four orders of magnitude lower and cannot be detected with a spectrometer/EMCCD. Instead, we use single photon avalanche detectors (SPAD, Micro Photon Devices PDM series) in combination with single-photon lock-in detection. Different non-linear contributions to the signal count rate are distinguished by modulating Stokes and probe beams at 50 kHz [Fig. 4(a)] as described in detail previously [22]. Using this lock-in detection, we can simultaneously record time delay scans of the 2pCARS, trCARS and trSERS signals (which were not accessible with spectral analysis) on each individual NPoM. To acquire sufficient photon counts at these low powers, time tracks are integrated for 60 minutes. We emphasise that this is the lowest laser intensity used for CARS experiments so far reported [10], which prevents damage to the nanostructures and molecules (demonstrated by stable count rates during long integration times, see Supplementary Information Fig. S4).

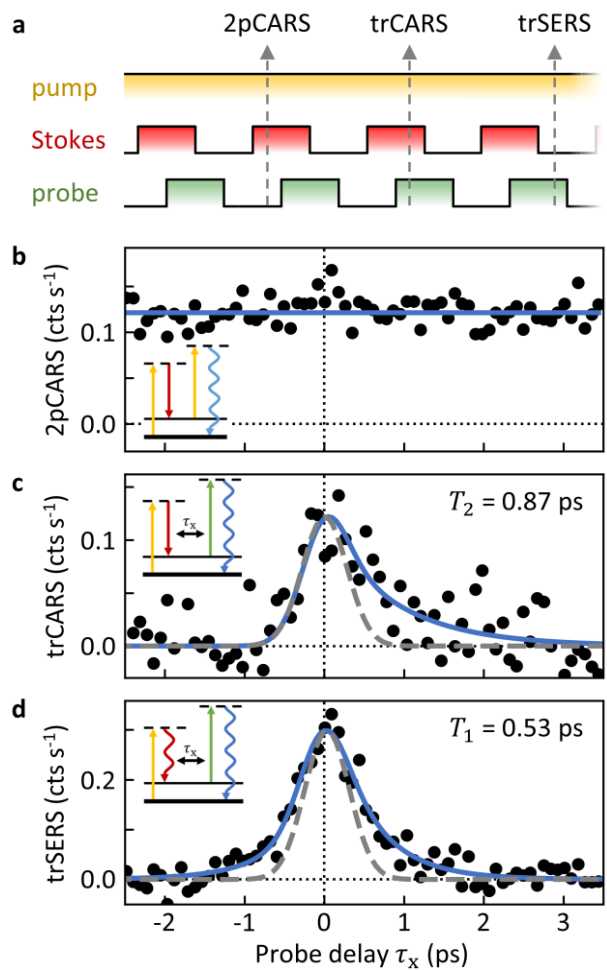

**Fig. 4 | Population and coherence decay from an individual NPoM. a,** Laser modulation scheme to identify different contributions to the signal with single-photon lock-in detection (50 kHz) [22]. **b,** 2-pulse CARS (2pCARS) signal remains unchanged during probe delay scan. **c,** Time-resolved CARS (trCARS) signal of an individual NPoM shows one-sided decay with probe delay giving coherence lifetime $T_2$. **d,** trSERS signal decays on both sides of probe delay scan with population lifetime $T_1$. Blue solid lines: exponential decay fit. Grey dashed lines: instrument response from a pure FWM sample (without molecular layer). Laser intensity 200 nW.µm$^{-2}$ per beam.

As expected, the 2pCARS is not modulated by sweeping the probe delay, showing the overall stability of the setup [Fig. 4(b)]. The trCARS signal at low laser power here shows a decay slightly longer than the instrument response (grey dashed) at positive probe pulse delays (probe arriving after pump-Stokes) [Fig. 4(c)]. The trCARS process probes the coherence of the system ($T_2$) and is influenced by both dephasing and decay of the vibrational population. For the NPoM shown here, we extract a coherence lifetime $T_2$ of 870 fs with an exponential fit convoluted by the instrument response [Fig. 4(c), blue]. In contrast, trSERS signals decay exponentially to both sides since both pump and probe pulses can excite vibrations via spontaneous Stokes scattering [Fig. 4(d)]. As an incoherent process, trSERS measures the population lifetime $T_1$ = 530 fs which is faster than the coherence decay for this NPoM. As the coherence decay rate $\Gamma_2 = 1/T_2$ is linked to the population decay rate $\Gamma_1 = 1/T_1$ via the pure dephasing rate $\Gamma_2^*$ through $\Gamma_2 = \frac{1}{2}\Gamma_1 + \Gamma_2^*$, [23] we can extract the pure dephasing time for each NPoM. In the particular case presented in Fig. 4 we observe a long dephasing time $T_2^*$ of 4.9 ps. We note from the time-frequency trade-offs required to resolve these vibrational lines that shorter laser pulses cannot be directly used for such measurements.

## V. Vibrational Energy Exchange

We perform trCARS experiments on 20 NPoMs to characterise the consistency of the observed decay rates. Fig. 5(a) shows the range of parameters extracted from probe delay scans on individual NPoMs. The pure dephasing rate $\Gamma_2^*$ varies considerably between NPoM constructs. To understand the origin of these deviations, we plot the decay rates against the maximum signal count rate at $\tau_x = 0$ ps [Fig. 5(b-d)]. A clear positive correlation between signal count rates and phonon decay rates is observed for both trCARS and trSERS. Particles with higher signal count rates show systematically faster population decay and dephasing. Other potential parameter correlations have been investigated (Supplementary Information Fig. S6), but no correlations are found.

Signal count rates vary strongly between NPoMs due to different light in- and outcoupling efficiencies governed by nanoparticle polydispersity and differences in contact facet size and shape [38,39]. Hence, the acquired signal count rate is closely connected to the laser intensity coupled into the gap and we can calculate the in-coupling efficiency $\eta_i$ for every NPoM $i$ as $\eta_i = S_i/\text{mean}(S_i)$ from each particle with signal count rate $S_i$. Using this efficiency, we can estimate the in-coupled laser intensity experienced by the NPoM as $I_l^{in} = \sqrt[3]{\eta_i} I_l$ (where the cubic root arises from $S_{CARS} \propto I_l^3$). Accordingly, higher signal count rates correspond to a higher laser intensity experienced by the system. Therefore, higher decay rates at higher count rates [Fig. 5(b-d)] are consistent with high-power experiments giving a strongly accelerated decay [Fig. 3(d)].

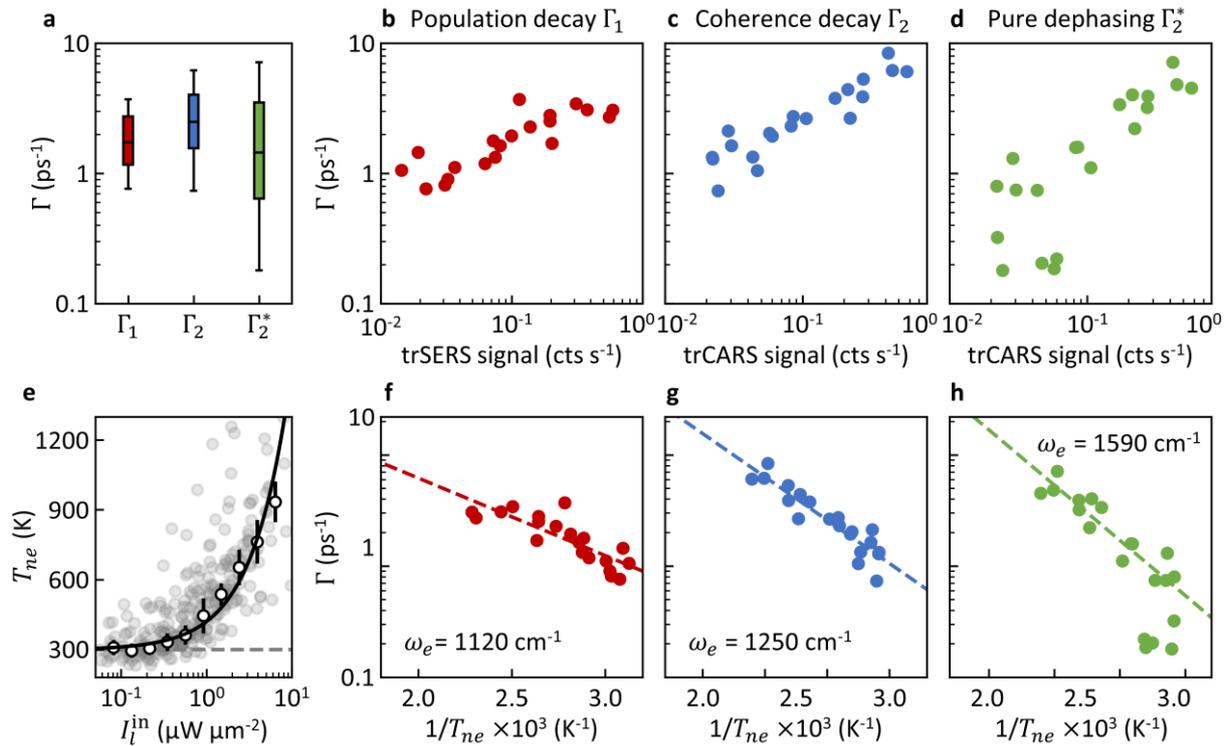

**Fig. 5 | Decay parameter analysis for many NPoMs. a,** Boxplots showing range of values for population decay $\Gamma_1$, coherence decay $\Gamma_2$ and pure dephasing rate $\Gamma_2^*$ from 20 NPoMs. **b-c,** Correlation of decay rate with amplitude of detected signal for (b) $\Gamma_1$, (c) $\Gamma_2$ and (d) $\Gamma_2^*$ for 20 NPoMs. For all three decay rates, a higher signal (and hence higher in-coupled laser intensity) is correlated with a faster decay. **e,** Non-equilibrium temperature $T_{ne}$ extracted from electronic anti-Stokes scattering *vs* in-coupled average laser intensity $I_l^{in}$ (at 785 nm) for >30 NPoMs. Fit to $T_{ne} = 300K + a. I_l^{in}$ (solid black line) giving temperature calibration for f-h. **f-h,** Decay rates vs. inverse temperature on many NPoMs for (f) $\Gamma_1$, (g) $\Gamma_2$ and (h) $\Gamma_2^*$ with fit of VEE model (dotted line). From the fit, the energy of the exchange mode $\omega_e$ can be determined.

The apparent temperature of the NPoM system under pulsed laser illumination can be independently analysed from the electronic Raman scattering (ERS) background. The non-equilibrium temperature $T_{ne}$ of the nanostructure at the pulse peak is estimated by fitting a Boltzmann distribution to the broadband anti-Stokes background of the pulsed SERS spectra [40] (Supplementary Information Fig. S5). An investigation of many NPoMs with a 785 nm laser (0.5 ps pulses at 80 MHz as for CARS experiments) reveals that the peak temperature increases with in-coupled laser intensity, reaching apparent non-equilibrium temperatures up to 1000 K above room temperature [Fig. 5(e)]. This compares to <1 K increases at typical CW laser powers used in SERS (at higher average powers). Hence, SE-CARS experiments even at lowest pulsed laser intensities probe the vibrational dynamics of a system transiently driven strongly out of thermal equilibrium by the laser pulses with the effective temperature highly dependent on each individual NPoM in-coupling efficiency. We thus use this ERS temperature to calibrate in-coupled laser intensities obtained in SE-CARS experiments to estimate NPoM non-equilibrium temperatures during each experiment [black line in Fig. 5(e)].

We observe $\ln(\Gamma) \propto 1/T_{ne}$ for both population decay $\Gamma_1$ and coherence decay $\Gamma_2$ (Fig. 5f-h). One previously-postulated explanation for this scaling is coupling of the resonantly excited phonon mode ($\omega_v$) to a lower-energy exchange mode at $\omega_e$ populated by the transient increase in temperature $T_{ne}$. [41,42] The vibrational energy exchange (VEE) model [43] describes this interaction of different phonon populations and proposes an increase of the decay rate $\Gamma$ due to the lower-energy mode as

$$\Gamma = \Gamma_0 + a \exp\left\{-\frac{\hbar\omega_e}{k_B T_{ne}}\right\} \ ,$$

where $\Gamma_0$ is the original decay rate and $a$ combines multiple parameters of the VEE model which cannot be disentangled with this setup. Fitting this model to the data allows us to estimate the energy of the exchange mode $\omega_e$ and thereby study the mechanism of vibrational energy decay and dephasing. For the population decay we find $\omega_e$ = 1120 $\pm$ 200 cm$^{-1}$, suggesting that the vibrational energy relaxes via intramolecular vibrational decay [44] to the 1080 cm$^{-1}$ mode of BPT (or nearby vibrations in this frequency range [45]). In contrast, the dephasing rate results in $\omega_e$ = 1590 $\pm$ 200 cm$^{-1}$ suggesting that dephasing is dominated by scattering with phonons of the mode $\omega_v$ itself. Investigations of a different molecule (triphenyl-4-thiol) and vibrational mode (1080 cm$^{-1}$ of BPT) confirm (Supplementary Information Fig. S7), within the margin of error of this model, that the population of the pumped vibration decays by intramolecular energy exchange to lower-energy modes while dephasing is driven by self-interaction of the pumped vibration. This intriguing result implies that fluctuations in the density of vibrations of this harmonic oscillator lead to nonlinear phase drift, potentially through anharmonicity of the vibrational potential. By contrast, the vibrational population decay is driven by a lower energy vibrational reservoir but can be tenfold faster than the dephasing suggesting a different coupling interaction.

Other mechanisms that lead to accelerated phonon decay, such as electron friction from the bond to the metal surface have been discussed in the literature but cannot reproduce the dependence on temperature found here [46]. Alternative models might involve ERS-induced non-thermal heating of the electrons in the metal, which then induce van-der-Waals-like dipole-dipole fluctuations on the vibrational dipoles, inducing their rapid decay. Further detailed theory based on these observations is therefore needed, and requires detailed models incorporating quantisation of multiple vibrations which is not yet available, but should be guided by these results.

This observation that phonon decay inside molecules is enhanced when the molecule is very close to a plasmonic metal surface is relevant for understanding many complex molecular phenomena. These include metal contact effects in molecular electronics, catalysis of chemical reactions at metal surfaces, heat transfer from molecules at metals for organic thermoelectrics, and other situations where the coupling of vibrational modes to electronic and energy transport is relevant.

## VI. Conclusions

In summary, we demonstrate how 2-pulse and 3-pulse optical nonlinearities resonant with molecular vibrations are strongly enhanced by plasmonic nanocavities. In the NPoM geometry, molecular CARS signals dominate over FWM as well as all electronic Raman contributions. Since the phonon lifetime is very sensitive to transient changes of the local electronic temperature, ultralow excitation powers are required to investigate vibrational decays. Using single-photon lock-in detection, we explore vibrational population decay and dephasing in BPT and show that vibrational energy exchange can account for the accelerated phonon decay of molecules on metallic surfaces. These ultrafast spectroscopy experiments on ~100 molecules open the way towards exploring single-molecule dynamics under ambient conditions and demonstrate a new optical probe for modified molecular dynamics under extreme optical field confinement.

## VII. Materials and Methods

### A. Sample preparation

Template-stripped gold substrates were prepared by thermal evaporation of 100 nm of gold on silicon, with glass substrates glued with UV glue Norland 81 (Thorlabs), UV-cured, and a fresh sample stripped off for each use. Self-assembled monolayers (SAMs) of biphenyl-4-thiol (BPT) were fabricated by soaking a template-stripped gold substrate in 1 mM BPT (Sigma-Aldrich) solution in 200-proof anhydrous ethanol overnight. After SAM formation, the sample was cleaned with ethanol and blown dry with nitrogen. Nanoparticle-on-mirror samples were prepared by depositing 50 µL of 80 nm gold nanoparticles (AuNPs, BBI Solutions) on BPT SAMs on template-stripped gold, which were then rinsed off with deionised water after 30 s and blown dry with nitrogen. Polymer-coated NPoM samples were prepared by spin coating a 100 nm layer of PMMA (950k molecular weight, 2wt% in anisole) with a first cycle at 500 rpm for 10 s, then a second at 2000 rpm for 45 s. Samples were then left to bake on a hotplate at 50°C for 5 min to remove residual anisole solvent.

### B. CARS spectroscopy

A Spectra-Physics MaiTai pump laser (80 MHz, 100 fs) drives an optical parametric oscillator (Spectra-Physics Inspire) to generate three pulse trains of different wavelengths. These are filtered by independently tuneable bandpass filters (PhotonETC LLTF contrast) with a spectral bandwidth of 1.5 nm and temporal width of 0.5 ps. Laser pulses are temporally aligned with motorized delay stages. Two of the three laser beams are modulated at 50 kHz with electro-optical modulators for lock-in detection.

NPoMs are located with an Olympus 0.9 NA, 100x darkfield objective in a custom-built, automated microscope. Laser pulses are focused on the sample and signal is collected in reflection by the same objective. Tuneable edgepass filters (Fianium Superchrome) remove the laser pulses and other background. Spectra are recorded with an Andor iDus CCD camera (DU416A) on an Andor Shamrock monochromator (SR-303i-B). Alternatively, a single-photon avalanche diode from Micro Photon Devices ($PD-100-CTD) was used to detect the signal.

Single-photon multiclock lock-in detection was realised by recording arrival timestamps of all photons at the detector along with timestamps from several reference signals (80 MHz laser repetition rate, 50 kHz laser modulation, 1 Hz delay stage sweep). Photon timestamping with picosecond accuracy was achieved using a field-programmable gate array board (Digilent Arty Z7) and specially designed software. Extracting the relative phase of the photon arrival compared to each reference signal allows to measure the periodic modulation of the signal during the reference period in time. This technique is described in more detail elsewhere [22].

**Acknowledgements**
We acknowledge support from European Research Council (ERC) under Horizon 2020 research and innovation programme THOR (Grant Agreement No. 829067), and PICOFORCE (Grant Agreement No. 883703), and UK EPSRC grants EP/L027151/1, EP/R020965/1. L.A.J. acknowledges support from the Cambridge Commonwealth, European & International Trust and EPSRC award 2275079. R.A. acknowledges support from the Rutherford Foundation of the Royal Society Te Apārangi of New Zealand, the Winton Programme for the Physics of Sustainability, and Trinity College, University of Cambridge. B.d.N. acknowledges support from Winton Programme for the Physics of Sustainability and the Royal Society (URF\R1\211162). N.S.M. acknowledges support from the German National Academy of Sciences Leopoldina.

**Author Contributions**
LAJ, WMD and JJB devised the experimental techniques and developed data analysis and theory. RA, BdN and NSM fabricated samples. All authors contributed to writing the manuscript.

**Conflicts of interest**
The authors declare no conflicts of interest.

**Corresponding author**
Correspondence to J.J.B. (jjb12@cam.ac.uk)

**Data availability**
Data for all figures can be found at DOI: (will be in Cambridge open data archive, at proof stage)

# Accelerated Molecular Vibrational Decay and Suppressed Electronic Nonlinearities in Plasmonic Cavities through Coherent Raman Scattering

# Supplementary Information

Lukas A. Jakob[1†], William M. Deacon[1†], Rakesh Arul[1], Bart de Nijs[1],
Niclas S. Mueller[1], Jeremy J. Baumberg[1*]

[1] Nanophotonics Centre, Cavendish Laboratory, University of Cambridge, Cambridge CB3 0HE, UK.
[†] Authors contributed equally; [*] jjb12@cam.ac.uk

## Contents

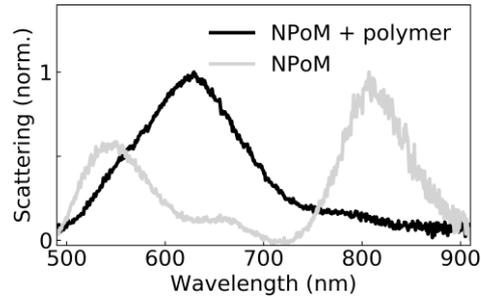

**Figure S1 | NPoM darkfield scattering.** Comparison of darkfield scattering spectra of individual nanoparticle-on-mirror (NPoM) constructs with (black) and without (grey) PMMA polymer coating. The polymer coating strongly enhances scattering around 650 nm benefitting CARS signal outcoupling.

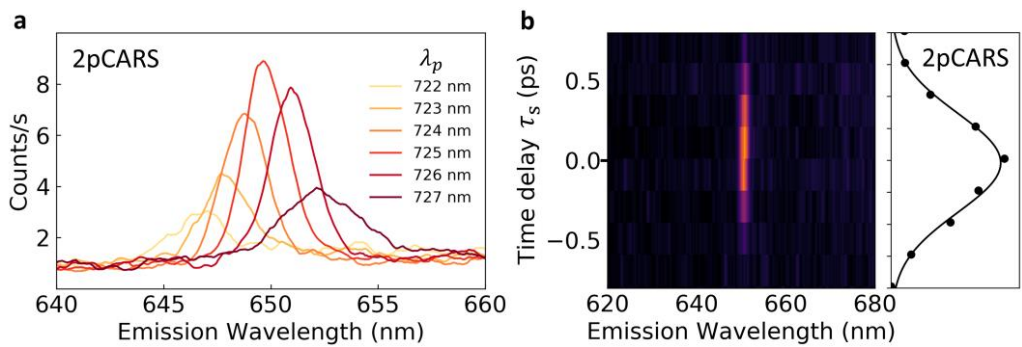

**Figure S2 | 2pCARS experiments. (a)** 2pCARS emission spectra (only pump and Stokes pulses on sample, $\tau_S$ = 0 ps) for different pump laser wavelength $\lambda_p$. Only when pump and Stokes pulses match the molecular resonance, strong CARS emission is observed. **(b)** Stokes pulse delay scan of 2pCARS spectra and integrated signal intensity (black line: Gaussian fit matching system response).

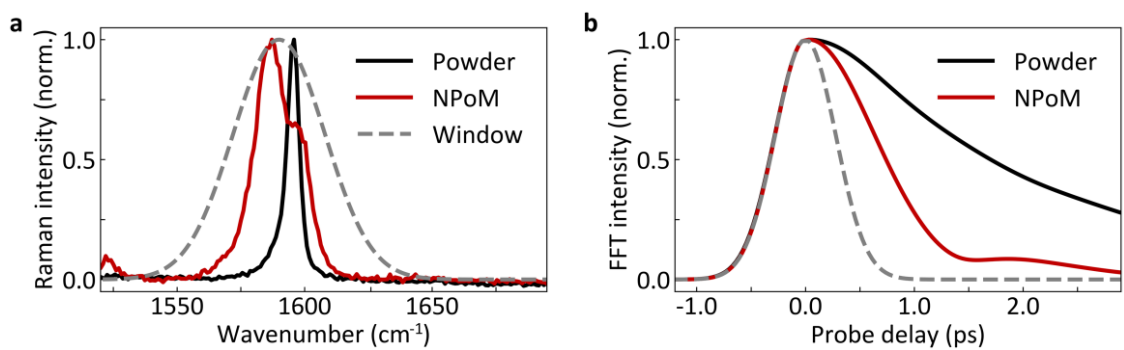

**Figure S3 | Fourier transformation of Raman spectra. (a)** 1585 cm$^{-1}$ mode of BPT in powder Raman (black) and NPoM SERS (red) with window function (grey, dotted) applied before Fourier transformation. The window function matches the spectral width (18 cm$^{-1}$) of pulsed lasers employed for trCARS experiments. Spectra recorded with a 785 nm continuous-wave laser. **(b)** Fourier transform of 1585 cm$^{-1}$ mode in BPT powder Raman (black) and NPoM SERS (red) spectra. The calculated time tracks simulate the expected trCARS response of the system. Fourier transform of window function (grey) represents the instrument response. Broadening of the Raman line in the NPoM leads to faster decay and the split mode causes a weak beating at 2 ps.

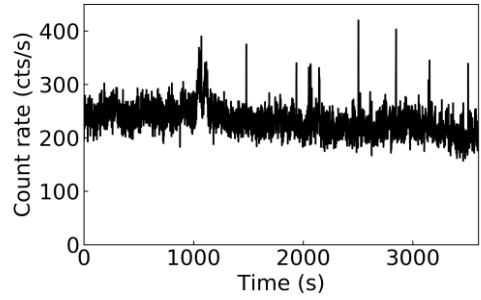

**Figure S4 | Sample stability under pulsed laser illumination.** Signal count rate on the single-photon detector during the trCARS measurement shown in the main text in Fig. 4. Short spikes in count rate are due to transient events such as picocavity formation, but no damage to nanostructure or molecules is observed at 200 nW.µm$^{-2}$ per beam during the integration time of 1 h. Dark count rate is <100 cts/s.

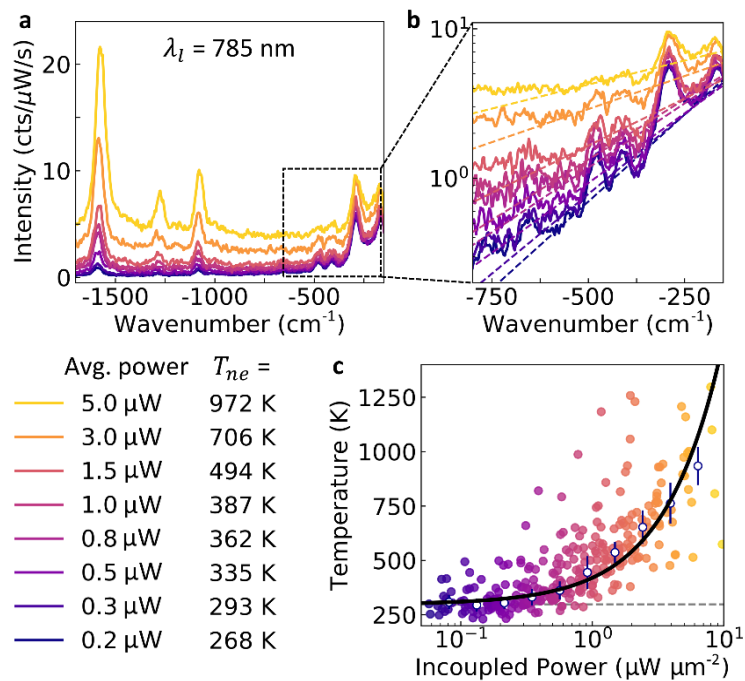

**Figure S5 | Electronic Raman scattering and temperature.** (a) Power-dependent (0.2 – 5.0 µW, colours) anti-Stokes spectra from BPT NPoM with 785 nm, 0.5 ps laser pulses. (b) Exponential fit to the low wavenumber background of the spectra to extract the apparent temperature of the electron gas under pulsed laser illumination. Particle temperatures strongly increase from room temperature to up to 1000 K at highest laser power (below damage threshold). (c) Non-equilibrium temperature vs. incoupling-corrected power for >30 NPoMs. Fit of $T_{ne}$ = 300K + $a. I_l^{in}$ (black curve) is used to calibrate incoupled power to temperature in main text Fig. 5.

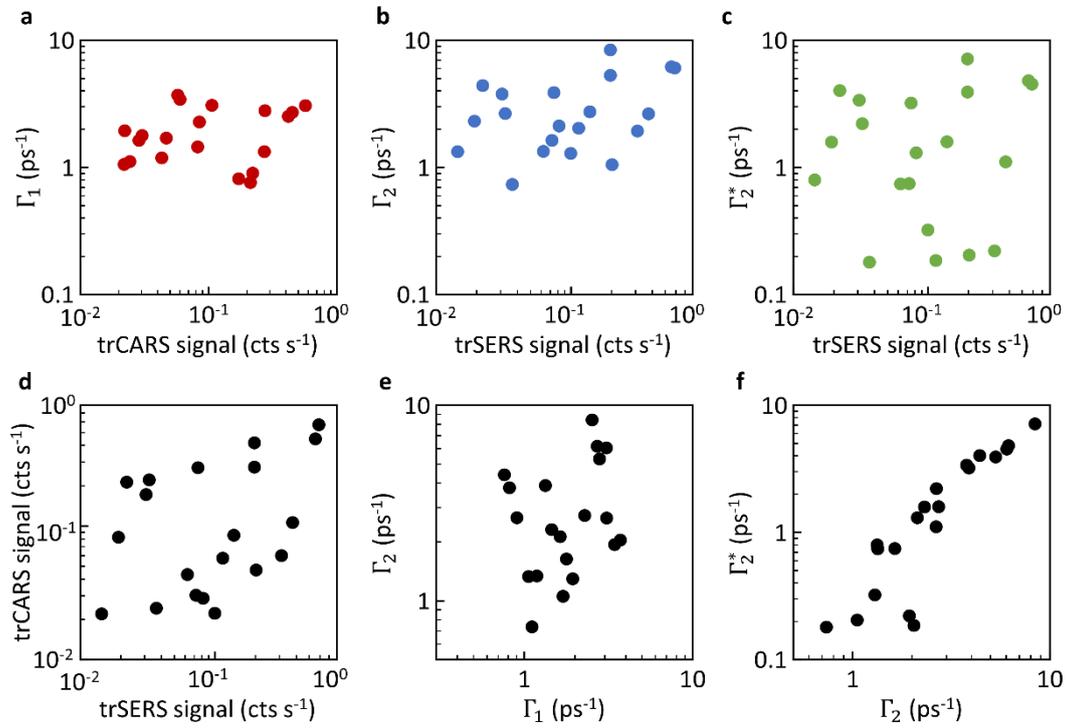

**Figure S6 | Exploration of other correlations in fit parameters of time tracks.** No significant trend is observed between **(a)** population decay rate vs trCARS signal, **(b)** coherence decay rate vs trSERS signal, **(c)** pure dephasing rate vs trSERS signal, **(d)** trCARS vs trSERS signal count rates, and **(e)** coherence vs population decay rates. This shows that trSERS and trCARS are separate physical processes with independent timescales and efficiencies. **(f)** Pure dephasing vs coherence decay rates show a strong correlation as these are both probed by trCARS and calculated from the same time track. Data from the same particles as shown in main text Fig. 5.

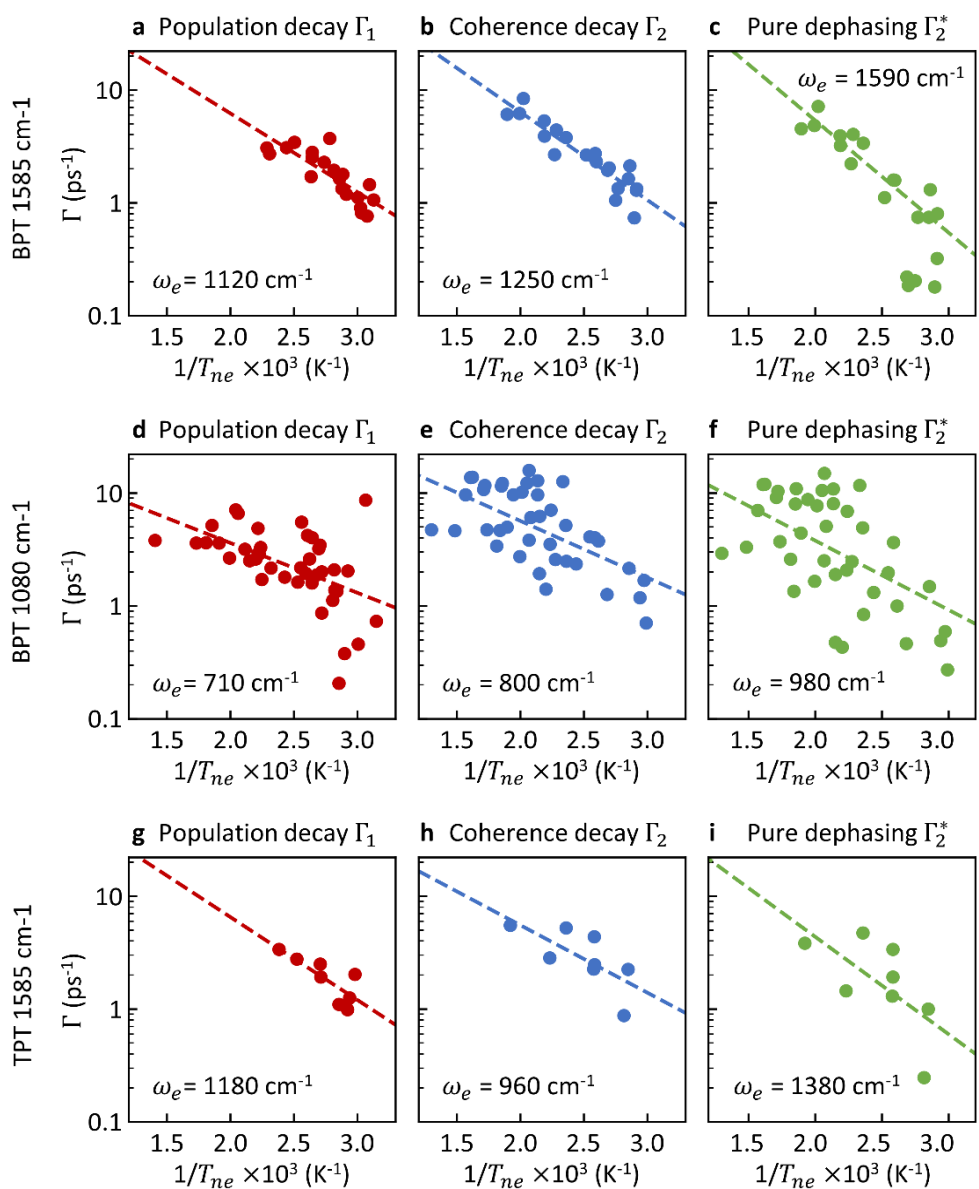

**Figure S7 | Decay rates of different vibrations and molecules.** Comparison of extracted decay rates vs non-equilibrium temperature of the biphenyl-4-thiol (BPT) **(a-c)** 1585 cm$^{-1}$ and **(d-f)** 1080 cm$^{-1}$ vibrations as well as the **(g-i)** 1585 cm$^{-1}$ mode of triphenyl-4-thiol (TPT). A fit of the vibrational energy exchange model identifies the frequency of the exchange mode $\omega_e$ (fit error $\pm$ 200 cm$^{-1}$). This reveals that the vibrational population for all molecules and modes decays via coupling to a lower energy vibration. Dephasing occurs for all vibrations with an exchange mode close to the original frequency.

5